# An additional source of gas ionization in the extended narrow line region of NGC1068.

**L.S. Nazarova**

Royal Greenwich Observatory, Madingley Rd.,Cambridge,CB3 0EZ,UK
Sternberg Astronomical Institute,Universitetskij prosp.13, Moscow, 119899, Russia

**Abstract.** The results of calculations of gas emission spectra with both central and extended sources of ionization have been compared to the ratio of line intensities observed in the extended narrow line region of NGC1068. The origin of an extended structure of anomalous strength in the $[OIII]\lambda 5007$ and $[NeV]\lambda 3425$ lines found by Evans & Dopita (1986) and Bergeron et al.(1989) could be due to an additional stellar source of gas ionization located at a distance 1-2 kpc from the nucleus.

**Key words:** galaxies:active – galaxies: individual(NGC1068) – galaxies: nuclei– galaxies: Seyfert– modelling: galaxies

## 1. Introduction.

The idea that the central region of Active Galactic Nuclei (AGN) can be separated into two regions, the broad- and narrow line regions (BLR and NLR), has been around for more than 25 years. Recently a third region of high excitation line emission has been discovered in some AGN. This region is more extended than the NLR and has been dubbed Extended Narrow Line Region (ENLR)(Unger et al. 1987; Bergeron & Durret 1987).

Research on ENLR spectra has led to the suggestion that the gas in the ENLR of active galaxies is illuminated by an anisotropic beam of ionizing UV- radiation and the emission from the broad line region could escape only in a cone. This idea has led to the so-called "Unified Scheme" of AGN according to which the difference between Seyfert 1 and Seyfert 2 galaxies is only due to the varying degree of obscuration and orientation (Antonucci & Miller 1985). The Unified Scheme can successfully solve the problem of "energy deficit" in AGN. This problem arises because the number of ionizing photons necessary for gas excitation in the ENLR is 10-100 times higher than that observed (Colina 1992).

NGC1068 is the first galaxy for which the ENLR has been discovered and intensively studied. The spectropo-larimetric observations (Antonucci & Miller 1985; Snijders et al. 1986; Miller & Goodrich 1990; Miller et al. 1991; Tran et al. 1992) show broad polarized FeII and Balmer emission lines in the nuclear spectrum of NGC1068. These results could be interpreted as the obscuration along the line of sight of non-thermal radiation from the central source by a disk or torus. Indeed, near-UV continuum (Pogge & De Robertis 1993) and X-ray images of NGC1068 (Wilson et al. 1992) show that the circumnuclear emission has a cone-like geometry and an elongated morphology in the direction of the radio jet (Haniff et al.

Pogge 1988,1989). Besides that the velocity field of ionized gas in the central region of NGC1068 reveals the presence of non-circular motion (Meaburn & Pedlar 1986; Baldwin et al. 1987; Bergeron et al. 1989) with a maximum deviation from circular motion also in the direction of PA= $32^o - 33^o$ (Pogge 1988; Jackson et al. 1993).

Inspection of the spectra of the ENLR in NGC1068 shows the existence of ionized gas emitting high-excitation $[NeV]\lambda3425$ and $[OIII]\lambda5007$ lines in a range of distances between $15''$ and $50''$ from the nucleus (Evans & Dopita 1986; Bergeron et al. 1989). The intensity ratios of lower ionization lines $I([OII]\lambda3727)/I(H_\alpha)$ and $I([NII]\lambda6583)/I(H_\alpha)$ are similar for HII regions and do not reflect a degree of ionization which might be expected from $[NeV]\lambda3425$ line strength (Evans & Dopita 1986). Furthermore the spectra of the ENLR show a low intensity of $[OIII]\lambda5007$ and $HeII\lambda4686$ lines compared to the intensity of $[NeV]\lambda3425$ line.

Evans and Dopita (1986) have modelled spectra of the ENLR of NGC1068 with two components of emission: HII regions and highly ionised low density gas. They also suggested that the continuum of the central source has a turn-on energy varying over the range 20-60 Ryd for the decrease in intensity of $HeII\lambda4686$ and $[OIII]\lambda5007$ lines compared to the intensity of $[NeV]\lambda3425$ line. If this photoelectric absorption mechanism of soft X-rays is assumed, then the comparison of the density given by the observed soft X-ray intensity and density, which is necessary for modelling the observed intensity of line emission, leads to the conclusion that the soft X-ray luminosity should be roughly 2 orders of magnitude larger than that observed (Bergeron et al. 1989). Consequently, this mechanism has definitely solved the "energy deficit" problem, but in the frame of the Unified Scheme only.

However, the suggestion that the excitation of the gas in the ENLR of NGC1068 is caused solely by source conflicts with the non-detection of hard X-rays in the 50kev and 100kev range (Jourdain et al. 1993; Maisack et al. 1994). This observation shows that if the Unified Scheme holds, the absorbing medium could be optically thick with a column density of at least $10^{24}$ cm$^{-2}$. This on the other hand conflicts with the observed peak of the IR emission in NGC1068 (Telesco & Harper 1980).

These considerations motivate a search for an alternative interpretation of the alignment effect in the ENLR of NGC1068. One such alternative concerns the existence of an additional extranuclear source of ionization: shocks or young stars.

In fact, the central region of NGC1068 has a ring of very luminous HII regions (Snijders et al. 1982; Bruhweiler et al. 1991) with a radius of approximately $13''$. The total bolometric luminosity for all observed starburst knots in the inner disk of NGC1068 is about $3.8\times10^{43}$ ergs s$^{-1}$ (Bruhweiler et al. 1991). Thus the UV flux ($\lambda$ 912) of the non-thermal radiation from the nucleus before photoelectric absorption may consist of about 20% - 30% of the UV flux from the starburst ring at a distance of 2 kpc. Evans and Dopita (1986) identified a blend of emission features due to CIII and CIV permitted lines near one of the starburst knots in the nuclear region. They interpreted this feature near $4660\mathring{A}$ as due to the presence of W-R stars of the carbon sequence (WC). Although the starburst ring in the nuclear region of NGC1068 has a circular structure, nevertheless we might expect that the combination of an anisotropic radiation from the nucleus together with radiation from hot stars could be a reason for a cone-like emission structure in some lines.

This paper presents the calculation of the emission spectrum for high excitation regions in NGC1068, taking into account both central and extended sources of ionization.The remainder of the paper is organised as follows: the code used for the calculations is described in Sec.2;

in Sec.3 and the conclusions in Sec.4.

## 2. The code.

The calculations were carried out using a new version of the code PHOTO. The first version of the code, developed for the study of planetary nebulae, is described in Golovayi & Mal'kov (1991). PHOTO calculates the emission spectra, given the incident continuum; the gas density as a function of distance from the central source; and the abundances of 10 elements: $H, He, C, N, O, Ne, Mg, Si, S, Ar$. The code allows the abundances to vary with the distance from the central source. However in this paper the line intensities has been calculated for solar abundances. A new feature of the present version is the calculation of spectral lines in different parts of a galaxy with both central and extended sources of ionizing radiation (thermal, power-law or a combination thereof). Photo- and collisional ionization, as well as radiative and dielectronic recombination (at high and low temperatures), and charge exchange between ions of heavy elements, and $H^0$ and $He^0$ atoms, are the processes which determine the intensities of the emission line. The equations of ionization-recombination (thermal and statistical) equilibrium are solved by standard methods, while heating and ionization of the gas by a diffuse radiation field have been calculated more accurately. The radiation produced by electron recombination to the ground state of $H^+$, $He^+$, $He^{++}$ and to the second $He^{++}$ level, as well as $L_\alpha$ transitions in HeI and HeII for a diffuse ionizing radiation field, have been taken into account. The diffuse radiation arriving at an arbitrary point from different directions, as a function of the angle between the center of the galaxy and the observer, is integrated. The extended source of gas ionization has been taken into consideration as an additional source in the diffuse radiation field.

velope into optically and geometrically thin layers of constant density. To obtain an accuracy of about 1% for the temperature in all layers, no more than 5 iterations are needed. The free parameters are: the abundance of the 10 elements, the incident continuum specified for energies between 480 Ryd and 0.5 Ryd, and the gas density as function of the distance from the centre. Furthermore we can take into account that the extended source of gas ionization is a function of the distance from the centre.

The comparison of the line intensities calculated with PHOTO and CLOUDY and other similar codes (Ferland 1991) showed that, for an identical parameter set, the predicted line intensities were generally in agreement to within 20%.

## 3. Model of high excitation gas in the ENLR of NGC1068.

The unusual spectra of extended high-excitation emission from some zones north-east of the nucleus NGC1068 have been modelled by Evans and Dopita (1986). Following them we also assume that the observed spectra of the extended emission region NGC1068 are due to the superposition of ordinary HII region spectra which are producing low-excitation line intensities, upon emission from the high-excitation gas.

However, in order to reduce the energy deficit problem it has been assumed that the luminosity of the central source in UV and soft X-ray is only 3.3 times larger than the observed luminosity in this range. The value $N_{rec}/N_{ion}$=3.3 gives the ratio of the number of recombination photons derived from hydrogen emission-line strengths to the number of photons which are necessary for gas excitation, and indicates that the ionizing continuum is anisotropic (Kinney et al. 1991).

The central non-thermal source of gas ionization ($F_\nu = F_o(\nu/\nu_o)^\alpha$) has a spectral index $\alpha=-1.6$, and the flux $F_o$

et al. 1991; Kriss et al. 1992). However the spectral index $\alpha=-1$ in the non-thermal continuum of the central source has been used for $\lambda \leq 228$ because the power-law of soft X-ray spectrum shows that the spectrum in this region could be decomposed into two components: a steep low-energy part ($\alpha=-3.5$) and a flat high-energy part ($\alpha \leq -1$) (Elvis & Lawrence 1987). Multi-aperture spectropolarimetric observations of NGC1068 have also led to the conclusion that the featureless continuum shape seems to be $F_\nu \propto \nu^{-0.88}$ (Antonucci et al. 1994). The gas densities are taken to be $N_o=1$ cm$^{-3}$ and the filling factor is about 0.01 for low density gas. An additional extended source due to hot stars of T=80000K and T=60000K located approximately at a distance of 1-2 kpc from the central source. The stellar atmosphere fluxes are taken from Clegg and Middlemass (1987). The densities of the hot stars are approximately $10^{-5}$ stars pc$^{-3}$ - $10^{-6}$ stars pc$^{-3}$.

### 3.1. Model with central power-law source.

At first glance the existence of the high-excitation $[NeV]\lambda 3425$ and $[OIII]\lambda 5007$ lines suggests that the excitation of the gas in the ENLR is due to the central power-law continuum. In fact in the excitation diagram $I([OIII]\lambda 5007)/I(H\beta)$ versus $I([NII]\lambda 6584)/I(H\alpha)$ most of the $[NeV]$ emission zones occupy positions similar to objects photoionized by power-law spectra (Bergeron et al. 1989).

In order to study changes of the line intensities with distance for the case of ionization by a power-law continuum, the variations of line intensity ratios with distance from the inner radius of the envelope, and different gas densities ($N_e=0.1$ cm$^{-3}$ and $N_e=1$ cm$^{-3}$) and different spectral indexes of the central non-thermal source ($\alpha=-1.5$ and $\alpha=-1$) have been calculated. The flux $F_o$ and the inner radius of the envelope have been varied, envelope are from $10^{-3}$ to $10^{-4}$.

According to our calculation with the sole central power-law source of the gas ionization the intensity ratio of lines I($[OIII]\lambda 5007$)/I($[NeV]\lambda 3425$) in the regions where line $[NeV]\lambda 3425$ radiated is about 6, although the observed intensity of these lines in ENLR of NGC1068 shows the ratio to be about 1-3.5 (Evans & Dopita 1986; Bergeron et al. 1989). Besides that, the observations also show a low intensity of $HeII\lambda 4686$ line which has a similar ionization energy as $[OIII]\lambda 5007$ line. The observed intensity ratio $I(HeII\lambda 4868)/I(NeV]\lambda 3425)$ equals 0.08-0.18. However our calculations for model with central power-law continuum shows this ratio to be about 0.2-0.4. Changing the electron density in our calculations, increasing the intensity or the spectral index of the non-thermal continuum displaces only the maximum intensity of lines from the central source. The calculations when the electron density is a function of the distance from the central non-thermal source, have also led to a shift of the ionizing front from the center but give a similar result for the lines intensity ratios I($[OIII]\lambda 5007$)/I($[NeV]\lambda 3425$) and $I(HeII\lambda 4868)/I(NeV]\lambda 3425)$.

### 3.2. Model with central and extended sources.

A reduction in the intensity of the lines $[OIII]\lambda 5007$ and $HeII\lambda 4868$ compared with the intensity of $[NeV]\lambda 3425$ line, is achieved by adding a stellar source. This source changes the population of the energy levels with the distance from the nucleus. The ions $O^{+2}$ move to the higher stages $O^{+4}$ and $O^{+5}$ and the ions $He^+$ to the $He^{++}$ stage. Similarly the ions $Ne^{+4}$ move to the stages $Ne^{+5}$ (Fig.1-3).

However the abundance of the $Ne^{+4}$ ions, which is caused by ionization of the ions $Ne^{+2}$ and $Ne^{+3}$ to the stages $Ne^{+4}$, shows a slow change with distance compared to the abundance of the $O^{+2}$ and $He^+$ ions. This leads to stronger radiation in the line $[NeV]\lambda 3425$ compared to that of $[OIII]\lambda 5007$ and $HeII\lambda 4686$.

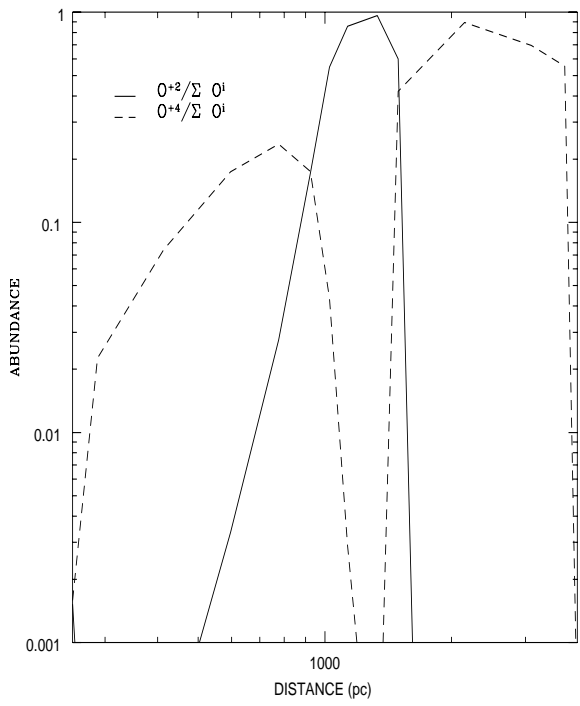

**Fig. 1.** The abundance of $O^{+2}$ (solid line) and $O^{+4}$ (dashed line) relative to total oxygen abundance as a function of the distance from the centre of NGC1068.

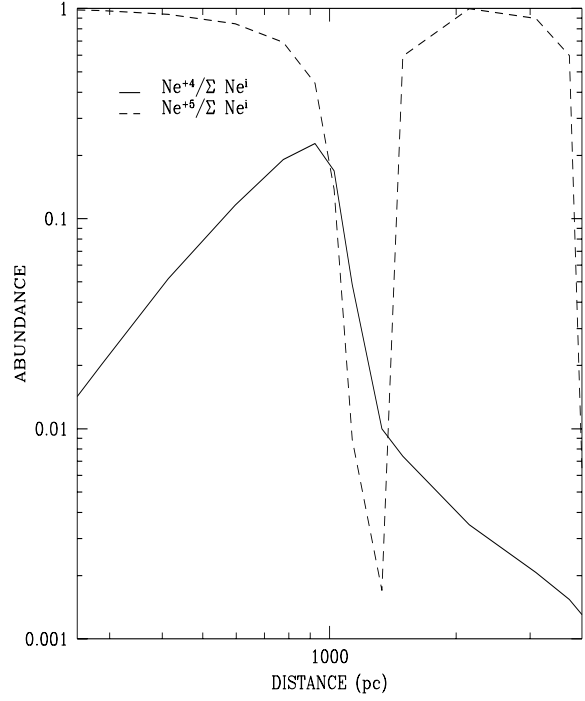

**Fig. 3.** The abundance of $Ne^{+4}$ (solid line) and $Ne^{+5}$ (dashed line) relative to total neon abundance as a function of the distance from the centre of NGC1068.

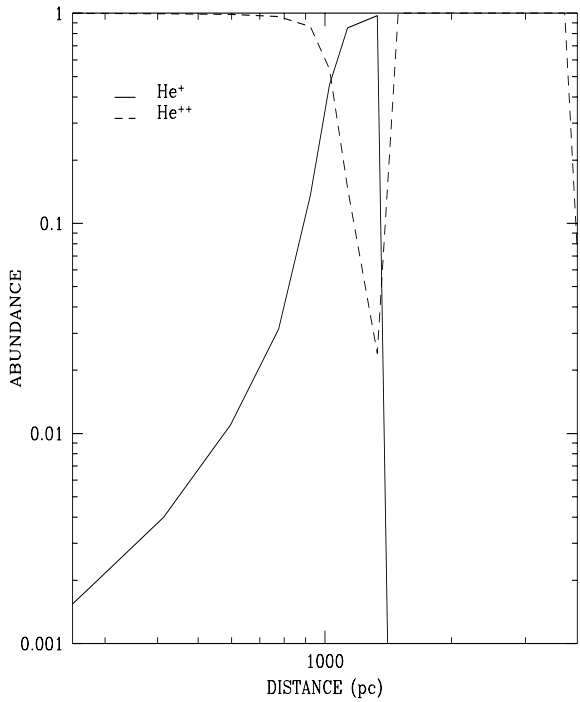

**Fig. 2.** The abundance of $He^+$ (solid line) and $He^{++}$ (dashed line) as a function of the distance from the centre of NGC1068.

Table 1 presents the characteristics of the models. The intensities of the mean observed spectra of the knots together with the theoretical line intensities (fluxes in the lines $[NeV]\lambda 3425 = 1$) for the 8 model spectra are presented in Table 2.

The line intensities of the observed extended emission spectra of NGC1068 shown in table 2 are derived from Bergeron et al. (1989) - (a) and from Evans & Dopita (1986) - (b). These two sets of observations almost coincide. The only line for which a significant discrepancy exists is [NeIII]$\lambda 3869$, which Evans & Dopita (1986) show to be much weaker. The 8 model spectra include the radiation from the regions of high and low excitation lines. Models 1-4 have been calculated with the extended stellar source with stars at a temperature of 80000K, while

**Table 1.** Model Parameters

| Models | 1 | 2 | 3 | 4 | 5 | 6 | 7 | 8 |
|---|---|---|---|---|---|---|---|---|
| Extended stellar source: | | | | | | | | |
| density of low ionization zone ($cm^{-3}$) | 20 | 30 | 50 | 100 | 20 | 30 | 50 | 100 |
| density of high ionization zone ($cm^{-3}$) | | | 1.00 | | | | 1.00 | |
| filling factor | | | 0.01 | | | | 0.01 | |
| temperature of stars (K) | | | 80000 | | | | 60000 | |
| density of stars (stars $pc^{-3}$) | | | $10^{-6}$ | | | | $10^{-5}$ | |
| Central source ($\alpha$): | | | | | | | | |
| $\lambda \leq 228$ | | | −1.0 | | | | −1.0 | |
| $\lambda \geq 228$ | | | −1.6 | | | | −1.6 | |

**Table 2.** Photoionization Models

| $\lambda$ | $Ion$ | (a) | (b) | 1 | 2 | 3 | 4 | 5 | 6 | 7 | 8 |
|---|---|---|---|---|---|---|---|---|---|---|---|
| 3425 | [NeV] | 1.0 | 1.00 | 1.00 | 1.00 | 1.00 | 1.00 | 1.00 | 1.00 | 1.00 | 1.00 |
| 3727 | [OII] | 1.09 − 3.78 | 1.47 − 2.25 | 2.22 | 2.09 | 2.02 | 1.85 | 1.18 | 1.13 | 1.1 | 1.00 |
| 3869 | [NeIII] | 0.38 − 1.05 | 0.09 − 0.20 | 0.16 | 0.18 | 0.18 | 0.19 | 0.06 | 0.05 | 0.05 | 0.05 |
| 4363 | [OIII] | ≤0.05-0.33 | 0.02 − 0.21 | 0.01 | 0.02 | 0.03 | 0.03 | 0.005 | 0.005 | 0.01 | 0.01 |
| 4686 | HeII | ≤0.13-0.18 | 0.08 − 0.18 | 0.05 | 0.05 | 0.05 | 0.05 | 0.03 | 0.03 | 0.03 | 0.03 |
| 4861 | H$\beta$ | 0.46 − 1.22 | 0.66 − 1.00 | 0.81 | 0.80 | 0.80 | 0.80 | 0.61 | 0.61 | 0.61 | 0.61 |
| 4959 | [OIII] | 0.64 − 1.13 | 0.35 − 0.53 | 0.65 | 0.68 | 0.76 | 0.87 | 0.21 | 0.21 | 0.23 | 0.28 |
| 5007 | [OIII] | 1.72 − 3.56 | 1.05 − 1.52 | 1.86 | 1.97 | 2.16 | 2.51 | 1.13 | 1.32 | 1.37 | 1.47 |
| 6300 | [OI] | ≤0.07-0.18 | | 0.15 | 0.15 | 0.13 | 0.12 | 0.08 | 0.08 | 0.08 | 0.08 |
| 6548 | [NII] | 0.21 − 0.79 | 0.42 − 0.64 | 0.24 | 0.24 | 0.24 | 0.19 | 0.16 | 0.15 | 0.15 | 0.12 |
| 6584 | [NII] | 0.95 − 2.55 | 1.20 − 1.82 | 0.72 | 0.68 | 0.64 | 0.44 | 0.48 | 0.45 | 0.41 | 0.38 |
| 6717 | [SII] | 0.28 − 0.83 | 0.3 − 0.47 | 0.75 | 0.70 | 0.64 | 0.51 | 0.55 | 0.52 | 0.45 | 0.42 |
| 6731 | [SII] | 0.12 − 0.47 | 0.21 − 0.34 | 0.69 | 0.67 | 0.45 | 0.41 | 0.40 | 0.32 | 0.32 | 0.31 |
| Fluxes (*) | | | | | | | | | | | |
| 3425 | [NeV] | 4.00 | | | | 3.92 | | | | 3.76 | |
| 4861 | H$\beta$ | 1.84 − 4.88 | | | | 3.13 | | | | 2.31 | |

(a)-observed dereddened line intensities from Bergeron et al. (1989)
(b)-observed line intensities from Evans & Dopita (1986)
(*)-Fluxes in the lines (units of $10^{-16}$ergs $cm^{-2}$ $s^{-1}$arcsec$^{-2}$)

stars at a temperature of 60000K have been used for the extended stellar source in models 5-8.

Inspection of table 2 illustrates the agreement between our theoretical models and the observed spectrum although the lines of [NII] are predicted to be weaker in our models. Evans & Dopita (1986) successfully modelled these lines with N/H abundance ratio 1.66 times higher than solar abundance. Possibly the solar abundances used in our models do not fit well the low ionization lines. The modelled line intensities ratio [NeIII]$\lambda$3869/[NeV]$\lambda$3425 coincides with this ratio observed by Evans & Dopita (1986) but differ from the results obtained by Bergeron et al.(1989). However the intensity of the line [NeV]$\lambda$3425 in the models coincides nearly with the observed value of $4\times 10^{-16}$erg $cm^{-2}$ $s^{-1}$ arcsec$^{-2}$ obtained by Bergeron et al. (1989).

The models agree better with the addition of a stellar source of ionization at a temperature of 80000K and electron densities for the low ionization regions about 30-50 $cm^{-3}$. The density of the stars located at an approximate distance 2 kpc from the nucleus is about $10^{-5}$ stars $pc^{-3}$ - $10^{-6}$ stars $pc^{-3}$. This gives the total luminosity of the

s$^{-1}$ which is close to the observed luminosity of the knots (Bruhweiler et al. 1991; Neff et al. 1994).

The possibility of a large stellar contribution from regions other than starburst knots has been discussed in several papers. The UV polarimetry data indicate an unpolarized starlight fraction near $\lambda$ 3300 to be about 0.3 even in the nuclear region of NGC1068 (4.3″ ×1.4″), which does not include the starburst knots (Antonucci et al. 1994). Wilson et al.(1992) have also suggested that the extended X-ray emission in NGC1068 could be produced by a starburst disk which consists of X-ray binaries. These stars may account for most or all of the hard spectrum emission from NGC1068 in the 2-10 kev band. In that model it is also no longer necessary to demand that the scattering cone in NGC1068 be optically thin below a few kev (Miller et al. 1991). However, this model is in conflict with the high degree of polarization (16%) in the UV continuum of NGC1068 (Code et al. 1993; Antonucci et al. 1994). Nevertheless one can observe that although the electron scattering model for UV and optical spectrum of NGC1068 gives a good fit to the nuclear flux it is unable to explain the near-infrared polarization of this galaxy (Young et al. 1994; Inglis et al. 1994).

Although there are many difficulties in explaining all the observations of NGC 1068 it seems possible to explain the emission in the line [NeV]$\lambda$3425 at the distance 14″ - 50″ from the nucleus of NGC1068 as a contribution of radiation from hot stars to non-thermal radiation from the central source. Of course there is strong evidence for anisotropic radiation in X-ray, UV, and radio regions in NGC1068. However many of these peculiarities have been discovered at a smaller distance from the nucleus (Ulvestad et al. 1987; Lynds et al. 1991; Evans et al. 1991; Kriss et al. 1992; Pogge & De Robertis 1993). The radio jets indicate emission on angular scales of 13″ (Ulvestad et al. 1987). The moderately strong CIII$\lambda$977 and NIII$\lambda$991 served only with the smallest circular aperture of 18″ (Kriss et al. 1992). The extended near-ultraviolet continuum emission also has a cone-like structure present inside the starburst ring (Pogge & De Robertis 1993). However only extended structures in some emission lines discussed in this paper show a cone-like structure at a large distance. Perhaps the propagation of anisotropic prominences near the nucleus at large distances is problematic, nevertheless the nuclear activity could of course be an indication of starburst processes far away from the nucleus.

## 4. Conclusions

The main difference between models of the spectra of the ENLR in previous theoretical work (Evans & Dopita 1986; Bergeron et al. 1989) and this paper resides on the assumption about the location of an ionizing source. In this paper, the high excitation gas of the ENLR of NGC1068 has been modelled with a power-law continuum, and with both central and extended sources of ionization. From the modelled spectra of the ENLR in NGC1068 we find the following:

1. The calculations for a power-law continuum with different intensities and shapes of the spectrum and for different gas densities show that the theoretical ratios of I([$OIII$]$\lambda$5007)/I([$NeV$]$\lambda$3425) and $I(HeII\lambda4868)$/I(NeV]$\lambda$3425) lines are higher than the one observed.

2. The calculations with both central and an extended ionizing source could explain the observed emission of the [NeV]$\lambda$3425 and (HeII$\lambda$4868) lines at a distance 14″ - 50″ from the nucleus.

3. An extended source of ionization in NGC1068 due to stars with temperature of 80000K and electron densities for the low ionization gas regions are about 30-50 cm$^{-3}$. The density of stars located at approximately 2 kpc from the nucleus ($10^{-5}$ stars pc$^{-3}$ - $10^{-6}$ stars pc$^{-3}$) gives a to $(2-3)\times 10^{43}$ergs s$^{-1}$.

4. The modelling of the ENLR spectra of NGC1068 with an additional stellar source in the ENLR also alleviates the "energy deficit" problem in AGN.

*Acknowledgements.* I am grateful to R.Terlevich for the initiation of this work and discussions. I want to thank S. Collin and A. Robinson and referee M.Durret for very fruitful comments and suggestion concerning the presentation of this paper. I wish also to thank the Royal Greenwich Observatory for their hospitality.